# NMR studies of the topological insulator $Bi_2Te_3$

A. O. Antonenko[1], E. V. Charnaya[1], D. Yu. Nefedov[1], D. Yu. Podorozhkin[1], A. V. Uskov[1], A. S. Bugaev[2], M. K. Lee[3], L. J. Chang[3], S. V. Naumov[4], Yu. A. Perevozchikova[4], V. V. Chistyakov[4], J. C. A. Huang[3], V. V. Marchenkov[4,5,6]

[1]Department of Physics, St. Petersburg State University, St. Petersburg, 198504 Russia
[2]Moscow Institute of Physics and Technology, Moscow, 141700 Russia
[3]National Cheng Kung University, Tainan, 70101 Taiwan
[4]M.N. Mikheev Institute of Metal Physics, UB of RAS, Ekaterinburg, 620137 Russia
[5]Ural Federal University, Ekaterinburg, 620002 Russia
[6]TU Wien Atominstitut, Vienna, 1020 Austria

$^{125}$Te NMR studies were carried out for the bismuth telluride, $Bi_2Te_3$, topological insulator in a wide range from room temperature down to 12.5 K. The measurements were made on a Bruker Avance 400 pulse spectrometer. The NMR spectra were collected for the mortar and pestle powder sample and for single-crystalline stacks with orientations $c \| \vec{B}$ and $c \perp \vec{B}$. The spectra at room temperature consisted of two lines which were attributed to the Te1 and Te2 positions. The parameters of the shift tensor were found from the powder spectrum. Temperature evolutions of the spectra for the powder sample and stack with $c \perp \vec{B}$ were self-consistent. The changes in the line positions with decreasing temperature were ascribed to a decrease in the Knight shift. The activation energy responsible for thermal activation of charge carriers was evaluated. The spectra for the stack with $c \| \vec{B}$ showed some particular behavior below 91 K. The Landau quantization on the topological insulator surface was suggested to respond for the anomalous NMR spectra in this orientation. Spin-lattice relaxation was studied for the powder sample and single-crystalline stack in both orientations at room temperature. The recovery of the longitudinal magnetization was single-exponential.

1. Introduction.

Topological insulators (TIs) are a new class of materials with a bulk electronic energy gap and topologically protected gapless surface states (see reviews [1,2]). In TIs the abstract space spanned by electronic wave functions has a nontrivial topology. The shrinkage of the energy gap to zero at the TI surface ensures a smooth transformation of the topology from the inner part of TIs to the outer space. The charge carriers at the surface behave like Dirac fermions with a spin locked to their momentum. Such surface states may have promising applications in quantum computation[3] and spintronics[4].

Bismuth telluride, $Bi_2Te_3$, belongs to 3-D TIs of the second generation. These crystals have been known as good thermoelectric materials and explored for a long time[5]. Theoretical and experimental studies showed that the nontrivial topology in $Bi_2Te_3$ and the emergence of surface states, as in other strong TIs, are associated with a band inversion in the presence of strong spin-orbit coupling[6-8]. Hence, the spin properties of electrons in the bulk and on the surface of $Bi_2Te_3$ are subject of increasing attention. Valuable information about the electronic spin polarization can be obtained by NMR. In semiconductors and metals the position and shape of NMR lines are influenced by coupling of nuclei to local magnetic fields induced by conduction electrons which are polarized by external magnetic field[9-11]. The resulting shift of the NMR frequency is called the Knight shift. The Knight shift for dipole nuclei which quadrupole moment equals zero, is summing up with the chemical shift[9]. In semiconductors the Knight shift generally shows strong temperature variations because of changes in the charge carrier density. Spin-lattice relaxation of nuclei in metals and semiconductors also depends on the e-spin polarization and hyperfine coupling with conduction electrons.

Only few recent works reported NMR studies of the second-generation 3-D TIs[12-17]. In particular, $^{125}$Te NMR studies of crystalline $Bi_2Te_3$ were carried out in Refs. [13,14,17]. The $^{125}$Te NMR spectrum and spin-lattice relaxation time were obtained in Ref. [13] for bismuth telluride powder. A single line was observed with a shift of about 500 ppm on the unified scale $\Xi$[18] at room temperature. The line position weakly depended on temperature within the range 150 to 400 K because of a large amount of native defects which produce charge carriers even at low temperature. Studies of $Bi_2Te_3$ nanoparticles by $^{125}$Te NMR at room temperature were presented in Ref. [14]. The lineshapes for nanoparticles of larger sizes were similar to those in $Bi_2Te_3$ powder. However, for smaller sizes a shoulder was observed at lower frequencies near -500 ppm which was treated as a signal from the nanoparticle surface. The $^{125}$Te NMR spectrum in single-crystalline $Bi_2Te_3$ consisted of two lines with shifts of 400 and -600 ppm[17]. The temperature dependence of the shift of the line with larger intensity (near 400 ppm at room temperature) was studied. The line moved to low frequencies with decreasing temperature following the thermal activation law.

To better understand the spin polarization in $Bi_2Te_3$, and generally those of 3-D TIs, it seems important to study the temperature evolution of $^{125}$Te NMR spectra in detail for powder as well as for single-crystalline samples. This should help to identify the spectrum components, separate the contribution of the Knight shift, and estimate the isotropic shift and anisotropy. In the present paper we report on results of $^{125}$Te NMR studies of the single crystalline $Bi_2Te_3$ and powder made of a bismuth telluride single crystal in a large temperature range below room temperature.

2. Samples and experiment.

The $Bi_2Te_3$ boule was grown by the Bridgman-Stockbarger method. It is known that the bismuth telluride crystals may have native defects which lead to a high amount of charge carriers at low temperature and weak dependence of conductivity on temperature. Then the perfection of the crystal lattice can be characterized by the ratio of resistance at 4.2 K and at room temperature. For the $Bi_2Te_3$ boule used in our measurements this ratio was equal to $\rho_{293K}/\rho_{4.2K} \approx 25$. This high ratio evidences the presence of only a small amount of native defects. The Hall effect monitored by a conventional technique within the range 4.2 to 80 K was negative corresponding to *n*-type conductivity.

Three plates were cut from the boule with their surfaces perpendicular to the crystallographic *c* axis. The thickness of the plates was near 0.2 mm and the surface area about 20 mm$^2$. The orientation of the plates was tested by x-ray diffraction. According to the x-ray diffraction two plates had several blocks which were rotated about the common *c* axis by an angle of about 10 degrees. To get a powder sample, a part of the bismuth telluride boule was ground by mortar and pestle.

The bismuth telluride crystals have a rhombohedral symmetry with space group $R\bar{3}m$ [19]. The unit cell comprises three formula units. The structure of $Bi_2Te_3$ consists of quintuple layers of atoms Te1-Bi-Te2-Bi-Te1 which are stacked along the *c*-axis (Fig. 1). The layers are bound by weak Van der Waals forces. The two tellurium sites, Te2 in the interior and Te1 in the exterior of the layers, are crystallographically nonequivalent. $^{125}$Te has spin ½ and therefore does not possess a quadrupole moment.

The $^{125}$Te NMR measurements were carried out in the temperature range from 12.5 K to room temperature on a Bruker Avance 400 pulse spectrometer. The operating frequency for $^{125}$Te was 126.26 MHz. A low-temperature static wide-line probehead with the cryosystem was used. The temperature stabilization was better than 0.5 K. To monitor the NMR spectra the spin echo pulse sequence $\pi/2-\tau-\pi$ with variable frequency offset was applied. The duration of the 90-degree pulse was in the range from 2.5 to 4 µs for all samples. Due to feeble signals the number of scans exceeded $2^{11}$.

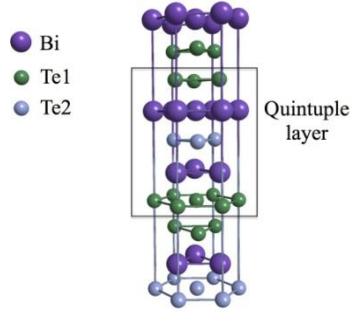

Fig. 1. Structure of the $Bi_2Te_3$ crystal.

Two kinds of spectra were collected. First, the intensities of spin echo at variable frequencies were found and then the envelope was plotted (envelope spectra). Second, the echo data at a stepped offset were summed up to get variable offset cumulative spectra. The temperature evolution of NMR spectra was studied for the powder sample and for three single crystalline plates stacked along their *c*-axis. The stack was arranged in magnetic field with $c \| \vec{B}$ and $c \perp \vec{B}$. The frequency of $^{125}$Te NMR was calibrated using the unified scale $\Xi^{18}$. To find the spin-lattice relaxation time the inversion recovery spin echo technique was used for the powder sample and the single-crystalline stack at $c \perp \vec{B}$ with the frequency offset at the two maxima of the spectra, and for the stack at $c \| \vec{B}$ with the frequency offset at the maximum of the more intensive line (see section 3). The spin-lattice relaxation time $T_1$ was measured only at room temperature because of a long time required for such measurements.

3. Results.

Fig. 2 (a) shows the $^{125}$Te envelope NMR spectrum at room temperature for the powder sample. The spectrum comprises of two quite separated lines with maxima at about 500 and -400 ppm. The temperature evolution of the envelope and cumulative spectra for the powder sample is shown in Fig. 3. Both lines move to low frequency rather identically with decreasing temperature. The high-frequency component of the spectra remains more intensive at all temperatures. The dependences of the line maxima on temperature are shown in Fig. 4. The line maxima were found using the envelope spectra.

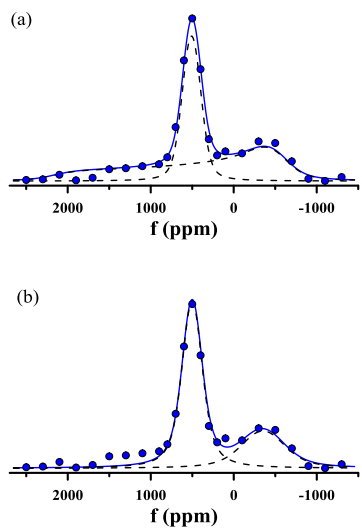

Fig. 2. Envelope $^{125}$Te spectrum for the Bi$_2$Te$_3$ powder at room temperature with two different fits as described in the text. Dash curves show the individual line fits and solid curves show their sum.

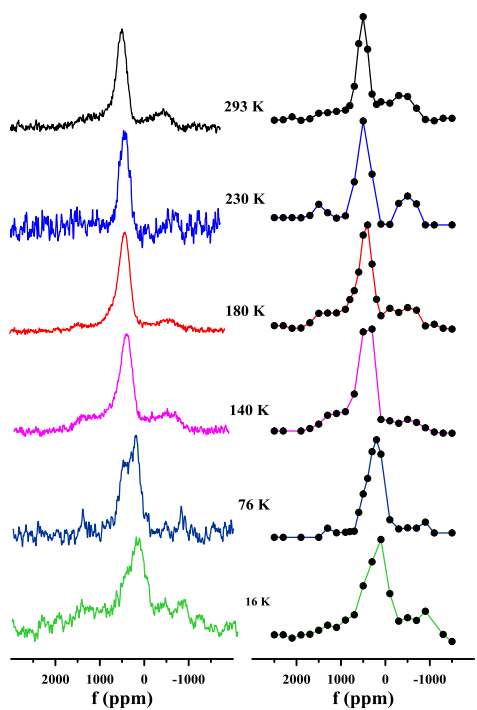

Fig. 3. Cumulative (left) and envelope (right) $^{125}$Te spectra for the Bi$_2$Te$_3$ powder at temperatures indicated on the panel. Solid lines on the envelope spectra are guides for the eye.

Fig. 5 shows the temperature evolution of the $^{125}$Te envelope and cumulative NMR spectra for the stack of single-crystal plates with the orientation $c \perp \vec{B}$. The spectra at room and lower temperatures are similar to the powder spectra. They also consist of two lines which shifts decrease with decreasing temperature similar to those for the powder spectra. The temperature evolution of the line maxima for the orientation $c \perp \vec{B}$ is shown in Fig. 6.

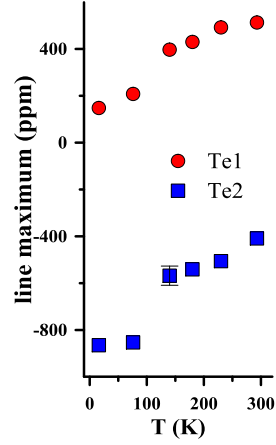

Fig. 4. Temperature dependences of the line maxima in the $^{125}$Te spectra for the $Bi_2Te_3$ powder. Circles and squares refer to high-frequency and low-frequency lines which were ascribed to Te1 and Te2, respectively. Error bars are shown only when they exceed the size of symbols.

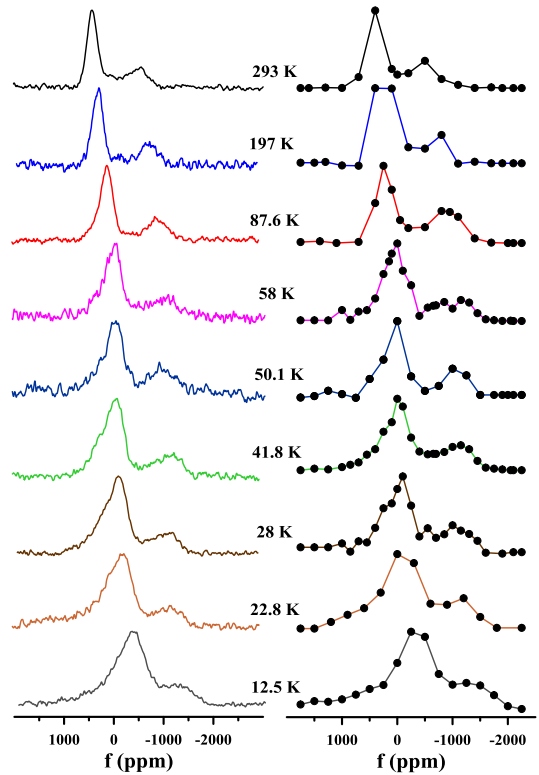

Fig. 5. Cumulative (left) and envelope (right) $^{125}$Te spectra for the stack of $Bi_2Te_3$ single-crystal plates with the orientation $c \perp \vec{B}$ at temperatures indicated on the panel. Solid lines on the envelope spectra are guides for the eye.

The spectra of the single-crystalline plates with orientation $c \| \vec{B}$ at different temperatures (Fig. 7) differ remarkably from those for the orientation $c \perp \vec{B}$ and from the powder spectra. At room temperature the more intensive line is near 800 ppm, while the second broad and weak line is now seen at higher frequency (at about 2200 ppm). The line centered near 800 ppm remains more intensive down to 132 K and does not shift noticeably.

At lower temperature the spectra for the orientation $c \parallel \vec{B}$ change drastically. At about 91 K the spectrum consists of two quite intensive lines which move to higher frequencies with decreasing temperature. At 16 K there is a very broad complex line with a maximum at about 1500 ppm.

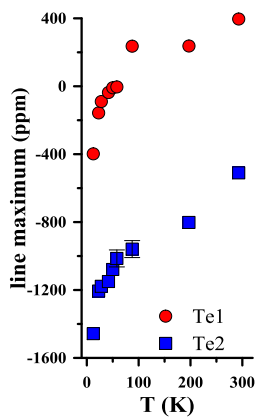

Fig. 6. Temperature dependences of the line maxima in the $^{125}$Te spectra for the stack of Bi$_2$Te$_3$ single-crystal plates with the orientation $c \perp \vec{B}$. Circles and squares refer to high-frequency and low-frequency lines which were ascribed to Te1 and Te2, respectively. Error bars are shown only when they exceed the size of symbols.

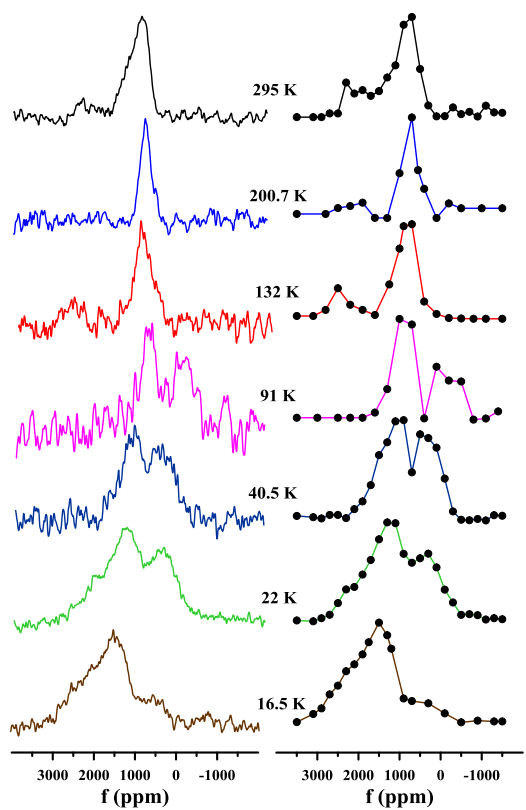

Fig. 7. Cumulative (left) and envelope (right) $^{125}$Te spectra for the stack of Bi$_2$Te$_3$ single-crystal plates with the orientation $c \parallel \vec{B}$ at temperatures indicated on the panel. Solid lines on the envelope spectra are guides for the eye.

The recovery of longitudinal magnetization at room temperature was single-exponential for all the samples. Some examples of $T_1$ measurements are shown in Fig. 8. The measurements were repeated several times and the average values of $T_1$ are listed in Table 1. For the stack with $c \| \vec{B}$ the relaxation measurements were carried out only for the intensive line at 800 ppm since the second line (with bigger shift) was weak. The relaxation rate for the $Bi_2Te_3$ powder and stack with $c \perp \vec{B}$ differed for two lines in the spectra, $T_1$ being noticeably shorter for the more intensive line. Relaxation was faster for the single-crystalline stack with $c \perp \vec{B}$ compared to the powder sample.

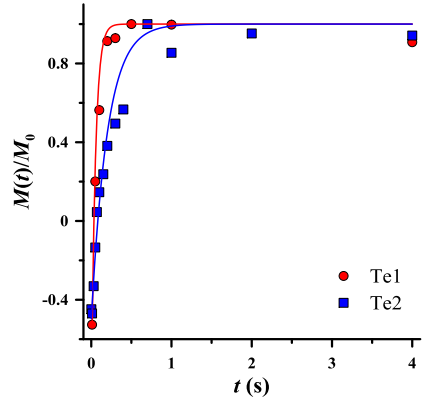

Fig. 8. The $^{125}$Te longitudinal magnetization recovery curves for the stack of $Bi_2Te_3$ single-crystal plates at the orientation $c \perp \vec{B}$. Circles and squares refer to high-frequency and low-frequency lines which were ascribed to Te1 and Te2, respectively. Solid lines show single-exponential fits.

Table 1. The $^{125}$Te spin-lattice relaxation times $T_1$ at room temperature for the $Bi_2Te_3$ powder and single-crystalline stack with $c \perp \vec{B}$ and $c \| \vec{B}$ for the lines associated with Te1 and Te2.

| sample | $T_1$ (ms), line Te1 | $T_1$ (ms), line Te2 |
|---|---|---|
| powder | $140 \pm 20$ | $400 \pm 50$ |
| stack $c \perp \vec{B}$ | $64 \pm 15$ | $230 \pm 30$ |
| stack $c \| \vec{B}$ | $130 \pm 20$ | |

4. Discussion.

Let us discuss first the powder spectrum at room temperature shown in Fig. 2. It consists of two separate lines with different longitudinal relaxation rates. The spectrum in Fig. 2 is in contrast with the single-line spectrum obtained earlier[13] for the $Bi_2Te_3$ powder. A two-component $^{125}$Te spectrum for $Bi_2Te_3$ was reported in Ref. [17], but only for a single-crystalline sample. As the structure of the $Bi_2Te_3$ topological insulator comprises two crystallographically non-equivalent tellurium ions, the observed lines can be attributed to Te1 and Te2. The amount of Te1 ions is twice as large as that of Te2. Therefore, the more intensive line should be associated with Te1 and the less intensive one with Te2. The NMR powder spectra due to the Knight shift and chemical shift are described by quite similar theoretical relationships[9]. Then the spectrum in Fig. 2 can be fitted using the total isotropic shift $\delta_{iso}$, anisotropy $\delta$, and asymmetry $\eta$ in the Haeberlen notation[20]. The spectrum deconvolution can be done in two different ways. The first way is illustrated in Fig. 2 (a). The more intensive line which corresponds to Te1 is rather symmetric and does not show features specific for an anisotropic shift tensor, i.e. the shift anisotropy is less or comparable to other broadening mechanisms. For the particular line shown in Fig. 2 the anisotropy should be less than 200 ppm. The isotropic shift for this line is 500 ppm. The second less intensive and strongly asymmetric line is characterized by an isotropic shift $\delta_{iso}$=380 ppm, anisotropy $\delta$=1800 ppm, and asymmetry $\eta$ close to 0. For such a deconvolution the ratio of the integral intensities of two lines is 0.56 which roughly agrees with the ratio of the Te2 and Te1 ions. An alternative way of deconvolution is shown in Fig. 2 (b). In this case both lines are about symmetric. One can only estimate their isotropic shifts which indeed coincide with the positions of their maxima. The ratio of the integral intensities of the two lines is 0.46 which also agrees with the ratio of the Te2 and Te1 ions. Both ways of fitting the powder spectrum seem to be feasible. The more appropriated fit can be chosen from other experiments.

The lines of the powder spectrum move to low frequencies with decreasing temperature. This trend is clear both in the cumulative as well as in envelope spectra. As can be seen in Fig. 3 the two lines move nearly identically. From data shown in Fig. 3 the temperature evolution of the isotropic shifts of the two lines can be found for two ways of spectrum deconvolution. As the isotropic shifts actually differ from the line positions only for the less intensive line when the first way of fitting was used, the temperature dependence of the isotropic shift is shown in Fig. 9 only for this case. To plot this dependence the envelope spectra were used.

The strong decrease of the isotropic shifts with decreasing temperature reflects the change of the Knight shift due to the decrease in the number of charge carriers[11,21]. Since the sample is of high quality according to resistance measurements (cf. section 2), the conduction charge density should decrease remarkably at low temperature. Then the shift at low temperature should be determined mainly by the chemical shift. As the spectrum shape at low temperature is about the same as at room temperature, we suggest that the anisotropy of the Knight shift is not stronger than that of the chemical shift.

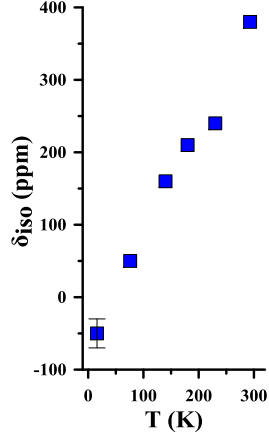

Fig. 9. Temperature dependences of the isotropic shift $\delta_{iso}$ for the line attributed to Te2 in the $^{125}$Te powder spectra for the first way of fitting. Error bars are shown only when they exceed the size of symbols.

The identical decrease in the line positions for two tellurium sites, Te1 and Te2, with decreasing temperature suggests that the Knight shifts for these sites are rather similar. The Knight shift for the Bi$_2$Te$_3$ was calculated in Ref. [22] for various densities of conduction electrons and holes. Pronounced difference between the Knight shifts for the two sites was obtained, the Te2 Knight shift being negative for electron and hole contributions. This does not agree with our experimental findings.

The Knight shift caused by charge carriers is proportional to the charge density divided by temperature[11]: $K_s \propto n/T$. Assuming thermal activation of the charge density for intrinsic semiconductors we have

$$K_s \propto n_0 \exp(-E_a/k_B T)/T, \qquad (1)$$

where $n_0$ is the charge density in the high-temperature limit, $E_a$ is the activation energy, and $k_B$ is the Boltzmann constant. According to (1), in order to get the activation energy, we should subtract the low-temperature magnitudes of shift from the total isotropic line shifts and plot $K_s T$ versus $1/T$. Such plots are shown in Fig. 10 (a) for the more intensive high-frequency line attributed to Te1 and in Fig. 10 (b,c) for the Te2 line in the powder spectra for the second and first models of fitting, respectively. The activation energy calculated from Fig. 10 (a-c) is the same within experimental accuracy and equal to $28\pm5$ meV. This activation energy is about three times smaller than the indirect energy gap obtained for Bi$_2$Te$_3$[23]. The reduction of the NMR shift with decreasing temperature normally evidences that the main contribution to the Knight shift comes from the conduction electrons[21]. In this case the activation energy can be written as $E_a = E_c - E_F$, where $E_c$ is the bottom of the conduction band and $E_F$ is the Fermi level. For an ideal intrinsic semiconductor the Fermi level lies in the middle of the band-gap, which leads to $E_a = E_g/2$. As for the Bi$_2$Te$_3$ powder under study the activation energy is smaller than $E_g/2$ we conclude that the Fermi level is moved to the conduction band. This corresponds to n-type conductivity which agrees with the negative sign of the Hall effect.

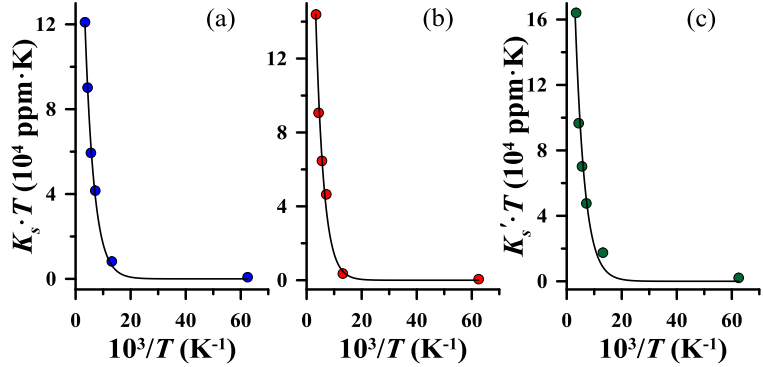

Fig. 10. Dependences of the product of the Knight shift and temperature on inversed temperature found from the powder spectra for the line attributed to the T1 site (a) and for the line attributed to the Te2 site assuming the second (b) and first (c) fits. $K_s$ is the difference between the values of line maxima in Fig. 4 at a particular temperature and in the low-temperature limit. $K'_s$ is the difference between $\delta_{iso}$ in Fig. 9 at a particular temperature and in the low-temperature limit.

The room temperature $^{125}$Te spectrum for the stack of three plated with $c \perp \vec{B}$ consists of two lines with positions 400 and -510 ppm which only slightly differ from the positions of lines in the powder spectrum. This agrees well with both ways of treating the powder spectrum suggested above. According to the second model the more and less intensive lines should centered within the widths of the lines in the powder spectrum. For the first model in the orientation $c \perp \vec{B}$ the less intensive line shift is determined by the $\delta_{xx}$ component of the shift tensor. Using $\delta_{iso}$ and $\delta$ found from the fit of the powder spectrum we have $\delta_{xx}$=-520 ppm which also agrees with the line position.

The temperature evolution of the $^{125}$Te spectra is rather similar to that of the powder spectra. Both lines move to lower frequency with decreasing temperature. The $K_s T$ versus 1/T plots for the two lines of the spectra are shown in Fig. 11. The calculated activation energies are very close to each other and differ by 1 meV, which is within error bars. The mean activation energy is $23 \pm 6$ meV. This value coincides within error bars with that estimated from the temperature evolution of the powder spectra. Therefore, the powder spectra and spectra for the single crystalline stack with $c \perp \vec{B}$ orientation are self-consistent.

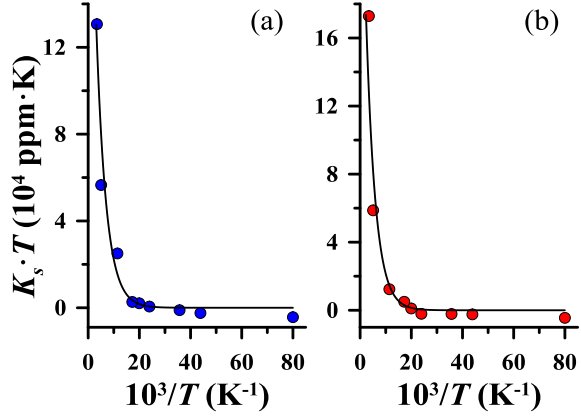

Fig. 11. Dependences of the product of the Knight shift and temperature on inversed temperature found from the spectra for the single-crystalline stack with $c \perp \vec{B}$. (a) and (b) refer to the lines attributed to Te1 and Te2, respectively. $K_s$ is the difference between the values of line maxima in Fig. 6 at a particular temperature and in the low-temperature limit.

The room-temperature spectrum for the stack with orientation $c \| \vec{B}$ also shows two lines. The position of the more intensive line slightly moves to high frequencies with a maximum at about 800 ppm. The difference between more intensive line positions at $c \| \vec{B}$ and $c \perp \vec{B}$ is of the order of the anisotropy $\delta$ estimated from the powder spectrum. Then, the more intensive lines for the two orientations of the single-crystalline plates are self-consistent and can be treated within the same model of the NMR line shift. The less intensive line for the plate orientation with $c \| \vec{B}$ is centered near 2200 ppm. The large shift of this line to high frequencies compared to the less intensive line in $c \perp \vec{B}$ orientation can be understood only within the framework of the first fitting procedure suggested above for the powder spectrum(see Fig. 2). The shift of this line at the $c \| \vec{B}$ orientation is determined by the $\delta_{zz}$ component of the shift tensor which equals 2180 ppm as can be found from the magnitudes of the isotropic shift $\delta_{iso}$ and anisotropy $\delta$ [20].

The temperature evolution of the spectra for the single-crystalline plates with $c \| \vec{B}$ orientation agree more or less with the evolution of the powder spectra and single-crystalline spectra with $c \perp \vec{B}$ orientation but only down to 132 K. Above this temperature the features of the spectra remain unchanged in general. However, at lower temperatures the spectra change drastically. At 91 K the less and more intensive lines exchange their positions. No regular drift of resonance lines to low frequencies with decreasing temperature can be seen and at 16 K the low-frequency line becomes very weak. This behavior is in complete contradiction with the powder spectra at low temperatures. Such an inconsistency might be related to the influence of the surface. In the orientation $c \| \vec{B}$ the surfaces of the plates are perpendicular to the external magnetic field. As a consequence the Landau quantization must be important for the surface states in this orientation. Due to strong spin-orbit coupling, the Landau quantization on the surface of TIs differs remarkably from that in two-dimensional conventional semiconductors[24] and leads to pronounced changes in the surface electron susceptibility. As the Knight shift of the NMR line for the surface of TIs remains roughly proportional to the electron susceptibility[25], the $^{125}$Te NMR line shift for the $c \| \vec{B}$ orientation of Bi$_2$Te$_3$ stack can show a particular behavior due to the surface contribution.

As was mentioned above and can be seen in Fig. 8, spin-lattice relaxation in the Bi$_2$Te$_3$ powder and single-crystalline samples at room temperature is single-exponential within experimental accuracy for both tellurium sites. For the high-frequency line ascribed to the Te1 site this result agrees with single-exponential recovery of the longitudinal magnetization for a single line which was observed in Ref. 13 for the Bi$_2$Te$_3$ powder. However, the relaxation time for this line in the powder spectrum (140 ms) is twice as long as the time (76 ms) reported in Ref. 13. The difference could be caused by different amount of lattice defects in the samples. For the mortar and pestle Bi$_2$Te$_3$ powder the longitudinal relaxation time was found in Ref. [14] to be equal to 133 ms in accordance with our measurements. The relaxation data for the second, low-frequency, line cannot be compared with previous results as this line was not observed in Ref. [13] and for the mortar and pestle Bi$_2$Te$_3$ powder in Ref. [14]. The much shorter relaxation time for the single-crystalline plates with $c \perp \vec{B}$ orientation may indicate a pronounced angular dependence of spin-lattice relaxation in Bi$_2$Te$_3$.

Conclusions.

The $^{125}$Te NMR spectra obtained for the Bi$_2$Te$_3$ powder showed two lines in the whole temperature range of measurements. The lines were attributed to the Te1 and Te2 sites. The spectrum was modelled assuming two sets of parameters of the shift tensor. Both models agree with the spectra for the single-crystalline Bi$_2$Te$_3$ stack with $c \perp \vec{B}$ orientation, however the high-temperature spectra for the stack orientation $c \| \vec{B}$ argue for the first way of fitting. The activation energy responsible for thermal activation of charge carriers was estimated. Its value indicates the displacement of the Fermi level to the conduction band in accordance with n-type conductivity in the bismuth telluride under study. The remarkable difference of the low-temperature spectra for the single-crystalline stack with the orientation $c \| \vec{B}$ from the spectra for the powder sample and stack with $c \perp \vec{B}$ could be related to the Landau quantization on the surface of TIs. Spin-lattice relaxation of $^{125}$Te was single-exponential at room temperature for both lines in the spectra of the bismuth telluride powder and stack with $c \perp \vec{B}$ and for the more intensive line in the spectrum for the stack with $c \| \vec{B}$.

**Acknowledgments**

The present work was supported by the Russian Foundation for Basic Research (grants No 14-02-92012 and 16-57-52009), by FASO of Russia (program "Spin" No. 01201463330), by the Government of the Russian Federation (state contract No. 02.A03.21.0006), and by the National Science Council of Taiwan. NMR and x-ray diffraction measurements were carried out at the Research park of St.-Petersburg State University. We thank Professor J. Haase (Leipzig University, Germany) for valuable comments and Professor H. W. Weber (TU Wien Atominstitut) for fruitful discussions and his careful reading the manuscript.

References

1. L. Fu and C.L. Kane, Phys. Rev. B **76**, 045302 (2007).
2. M. Z. Hasan and C. L. Kane, Rev. Mod. Phys. **82**, 3045 (2010).
3. N. Read, Phys. Today **65**, 38 (2012).
4. D. Pesin and A. H. MacDonald, Nat. Mater. **11**, 409 (2012).
5. S. K. Mishra, S. Satpathy, and O. Jepsen, J. Phys. Cond. Matter **9**, 461 (1997).
6. Y. L. Chen, J. G. Analytis, J. H. Lu, Z. K. Liu, S. K. Mo, X. L. Qi, H. J. Zhang, D. H. Lu, X. Dai, Z. Fang, S. C. Zhang, I. R. Fisher, Z. Hussain, and Z. X. Shen, Science **325**, 178, 2009.
7. D. Hsieh, Y. Xia, D. Qian, L. Wray, F. Meier, J. H. Dil, J. Osterwalder, L. Patthey, A. V. Fedorov, H. Lin, A. Bansil, D. Grauer, Y. S. Hor, R. J. Cava, and M.Z. Hasan, Phys. Rev. Lett. **103**, 146401 (2009).
8. Y. Xia, D. Qian, D. Hsieh, L. Wray, A. Pal, H. Lin, A. Bansil, D. Grauer, Y. S. Hor, R. J. Cava, and M.Z. Hasan, Nat. Phys. **5**, 398 (2009).
9. A. Abragam, Principles of Nuclear Magnetism (Oxford University Press, Oxford, 1985).
10. J. Winter, Magnetic Resonance in Metals (Clarendon Press, Oxford, 1971).
11. H. Selbach, O. Kanert, and D. Wolf, Phys. Rev. B **19**, 4435 (1979).
12. S. Mukhopadhyay, S. Krämer, H. Mayaffre, H. F. Legg, M. Orlita, C. Berthier, M. Horvatić, G. Martinez, M. Potemski, B. A. Piot, A. Materna, G. Strzelecka, and A. Hruban, Phys. Rev. B **91**, 081105 (2015).
13. R. E. Taylor, B. Leung, M. P. Lake, and L. S. Bouchard, J. Phys. Chem. C **116**, 17300 (2012).
14. D. Koumoulis, T. C. Chasapis, R. E. Taylor, M. P. Lake, D. King, N. N. Jarenwattananon, G. A. Fiete, M. G. Kanatzidis, and L. S. Bouchard, Phys. Rev. Lett. **110**, 026602 (2013).
15. B.-L. Young, Z.-Y. Lai, Z. Xu, A. Yang, G.D. Gu, Z.-H. Pan, T. Valla, G. J. Shu, R. Sankar, and F. C. Chou, Phys. Rev. B **86**, 075137 (2012).
16. N. M. Georgieva, D. Rybicki, R. Guehne, G. V. M. Williams, S. V. Chong, K. Kadowaki, I. Garate, and J. Haase, Phys. Rev. B **93**, 195120 (2016).
17. D. Yu. Podorozhkin, E. V. Charnaya, A. Antonenko, R. Mukhamad'yarov, V. V. Marchenkov, S. V. Naumov, J. C. A. Huang, H. W. Weber, and A. S. Bugaev, Phys. Solid State **57**, 1741 (2015).
18. R. K. Harris, E. D. Becker, S. M. C. De Menezes, R. Goodfellow, and P. Granger, Pure Appl. Chem. **73**, 1795 (2001).
19. W. Wong-Ng, H. Joress, J. Martin, P. Y. Zavalij, Y. Yan, and J. Yang, Appl. Phys. Lett. **100**, 082107 (2012).
20. M Mehring, Principles of High Resolution NMR in Solids (Springer Verlag, Berlin, 1983).
21. J. Y. Leloup, B. Sapoval, and G. Martinez, Phys. Rev. B **7**, 5276 (1973).
22. S. Boutin, J. Ramírez-Ruiz, and I. Garate, Phys. Rev. B **94**, 115204 (2016).
23. B. Yu. Yavorsky, N. F. Hinsche, I. Mertig, and P. Zahn, Phys. Rev. B **84**, 165208 (2011).
24. Z. Wang, Z.-G. Fu, S.-X. Wang, and P. Zhang, Phys. Rev. B **82**, 085429 (2010).
25. M. M. Vazifeh and M. Franz, Phys. Rev. B **86**, 045451 (2012).